\documentclass[12pt,a4paper]{article}
\usepackage{jheppub_kim}
 \usepackage{slashed}
\usepackage{longtable}
\usepackage{epsfig}
\usepackage{amssymb,amsmath}
\usepackage{amsthm}
\usepackage{graphicx}
 \usepackage{graphics}
\usepackage[hang,nooneline,scriptsize]{subfigure}
\usepackage{subfigure}
\usepackage{array}
\begin{document}
  
\title{Geodesic Motions near an improved Schwarzschild black hole }
 
\author[] {Surajit Mandal}
   
\affiliation[]{Department of Physics, Jadavpur University, Kolkata, West Bengal 700032, India }
	
 \emailAdd{surajitmandalju@gmail.com} 
\abstract{In this paper, we studied the geodesics of timelike and null like particles near an improved Schwarzschild black hole. The lapse function has been plotted and was found that only one horizon is possible. The equation of motion and effective potential of test particle have been calculated. This equation has an importance in studying the radial free fall and also in studying the stability of radial orbits (trajectories). The energy and angular momentum have also  been calculated to analysis the cicrular motion and stability of circular orbits. Moreover, Innermost stable circular orbit radius has determined. To get
a deeper insight of the nature of these trajectories, we have studied the timelike and null geodesics with the help of the dynamical systems approach. This analysis help us to determine the stability as well as the fixed point of phase space trajectories.

}	
	\maketitle

\section{Introduction}
Schwarzschild solution is necessarily a vacuum solution of Einstein's field equations, first developed by Schwarzschild in 1916 \cite{g1}. This vacuum solution has importance in general relativity to describe a black hole which is spherically symmetric in nature. The solution of improved Schwarzschild was first derived by Bonanno and Reuter \cite{g85}. Nowadays it is very challenging issues from theoretical prespective when quantum effect of geometry are taken into consideration. It is believed that those quantum effects will make a key role in the evaporation process of black hole(Planck size) as well as during the late stages of the gravitational collapse. According to semiclassical point of view, at a temperature a black hole emits Hawking radiation which is inversely proportional to the mass of the black hole. In addition to this process of Hawking radiation having continuous energy, a negative energy flux through the horizon can produced. As a result, mass of the black hole will be lowered and the temperature will be increased. It is yet unknown to us whether this process will continues until the whole mass of the black hole is converted to radiation or whether it will stops when the temperature of radiative black hole reaches to the Planck temperature limit where the semiclassical scenario are likely to break down.

One could speculate that the evaporation process comes to an end where the Schwarzschild radius is very close to the Planck length where the semiclassical results are no longer valid. Consequently the final state of the Hawking evaporation could be some kind of “cold” remnant has a mass not too far above the Planck mass. It is evident that the problem of the final state might be addressed within the framework of quantum gravity theory. The matter field is quantize only in semiclassical derivation of the Hawking temperature and it helps us to consider the spacetime metric as a fixed classical background. Moreover, quantum fluctuations of the metric has an important
role to inquire the black holes with a radius close to Planck length. The perturbative quantization of Einstein gravity has a little importance here since
it provides a non-renormalizable theory. The fundamental theory of quantum gravity such as string theory, loop quantum gravity, etc., do not
gives us a satisfactory answer yet \cite{h3}.  In order to study quantum effects in Schwarzschild spacetime, the idea of the Wilsonian renormalization group \cite{h4} comes into the picture and this essentially leads to an improved Schwarzschild black hole. Also, RG improvement of coupling effect on Schwarzschild black hole using action improvement has been investigated by R. Moti and A. Shojai \cite{g2, g3}.

Geodesic motion is not only a problem of theoretical interest. This has far-reaching consequences for the observations of these objects, as the motion of light and gravitational waves on geodesic trajectories have consequences for the observation of them, via shadows and mergers. In order to check the viability of geodesic motion near black hole, it is important to study the geodesic motion in four-dimensions as well as in higher-dimensions. For n-dimensional rotating black holes the equations comes from geodesic study are completely integrable, have studied by Don N. Page and others \cite{g15}. Geodesic study in Schwarzschild–Ads space–time can ben found in \cite{g11}. The geodesic structure for the Schwarzschild-Ads black hole has been studied by Norman Cruz and others \cite{g12}. The motion of test particle around a black hole in a braneworld can be found in \cite{g13}. Moreover, the motion of test particle in different spacetimes was found in \cite{g19, g20, g21, g22, g23}. Stuchlik and Calvani have studied on the effective potentials in case of radial null geodesics in
Kerr-de sitter and RN-de sitter spacetimes \cite{g14}. The study on bound orbits has been found in \cite{g16, g17}. Hackmann and others \cite{g18} have been investigated the behavior of timelike particles in four-dimensions. Recently, Wajiha Javed and others \cite{g4} have studied on deflection angle and greybody bound of an improved Schwarzschild black hole. Their analysis did not address the geodesic study around this black hole. In this paper we tried to do this analysis using effective Newtonian potential method. This paper also includes dynamical approach to understand the phase-space trajectories near this black hole.

The paper is organized in six parts. In Section, \ref{sec2}, the line element and the lapse function has been defined and the lapse function has been plotted to understand the nature of event horizon. In Section \ref{sec3}, the equations for the geodesics motion in 4-dimension around the black hole are found. In Section \ref{sec4}, to
understand the nature of orbits of test particles the method of effective Newtonian orbit calculation has been used. The equation of motion of the test particle of unit mass is estimated and the test particle moves under effective potential with a given energy and angular momentum has found to depend on the variuos parameter like free parameter, fixed parameter, mass of the black hole and also on the radial coordinate. This equation helps us to study radial motion of test particle and the corresponding stability of radial orbits. To do analysis the circular motion of test particles and the stability of circular orbits, we have used the method of transformation and determined the energy and angular momentum of massive particles. The radius of the innermost stable circular orbit (ISCO) of timelike (massive) particles is completely depends on their angular momentum and fixed parameter for a vanishing free parameter. To get a deeper insight of the nature of these trajectories, we have studied the timelike and null geodesics with the help of the dynamical systems approach in Section \ref{sec13}. The fixed points are found and the nature of the phase space trajectories have been studied. Finally we make the summary and conclusions in Section \ref{sec17}.


\section{ Preliminaries}\label{sec2}
The line element for an improved Schwarzschild black hole spacetime is defined as,
\begin{equation}\label{1}
ds^2=-f(r)dt^2+\frac{dr^2}{f(r)}+r^2d\Omega_{2}^2
\end{equation}
where r is the radial coordinate and
\begin{equation}\label{2}
d\Omega_{2}^2=d\theta^2+Sin^2 \theta d\phi^2
\end{equation}
Here, $d\Omega_{2}^2$ denotes the line element of unit two dimensional sphere. For this static and the spherically symmetric spacetime, the value of lapse function f(r) is defined as, \cite{g85, g86}
\begin{equation}\label{3}
f(r)=1-\frac{2Mr^2}{r^3+\tilde \omega(r+\psi M)}
\end{equation}
after simplifying the function (\ref{3}) takes the form as \cite{g4}
\begin{equation}\label{4}
f(r)=1-\frac{\tilde \omega^2}{r^4}-\frac{2\tilde \omega^2\psi M}{r^5}-\frac{2M}{r}+\frac{2M\tilde \omega}{r^3}=\frac{\Delta}{r^5}
\end{equation}
here M is the geometric mass of black hole , r is the radial coordinate, $\psi$ denotes free parameter and $\tilde \omega$ is fixed parameter. Fixed parameter $\tilde\omega$ and free parameter $\psi$ can take the value \cite{h2} 
$$\tilde\omega=\frac{118}{15\pi}$$ and $$\psi\ge 0$$
For a given M, $\tilde\omega$ and $\psi$, the horizon function $\Delta$ depends only on the
radial coordinate r. Pluggingg the value of f(r) in equation (\ref{1}), as a result we get the following
\begin{equation}\label{5}
ds^2=-\frac{\Delta}{r^5}dt^2+\frac{r^5}{\Delta}dr^2+r^2d\theta^2+r^2Sin^2 \theta d\phi^2
\end{equation}
The trajectories of the particle for this static solution are more complicated than the Schwarzschild case. In order to make the analysis simpler, the geodesics of this metrics are generally studied in terms
of an effective Newtonian potential. The behavior of the
real singularity at r = 0, depends on the parameter $\tilde \omega$ and $\psi$, which we assume in such a manner that there won't be any space-like naked singularity in the spacetime. A comparison study of the effective potentials for both the Newtonian and the Schwarzschild case has been discussed in \cite{g28}. We can analyze the positions of the horizons of this black hole by making $g^{rr}=0$, i.e
\begin{equation}\label{6}
f(r)=\frac{\Delta}{r^5}=0
\end{equation}
where $\Delta$ = $r^5-2Mr^4+2M\tilde \omega r^2-\tilde\omega^2 r-2\tilde\omega^2\psi M$

\begin{figure}[h!]
\begin{center} 
 $\begin{array}{cccc}
\subfigure[]{\includegraphics[width=0.5\linewidth]{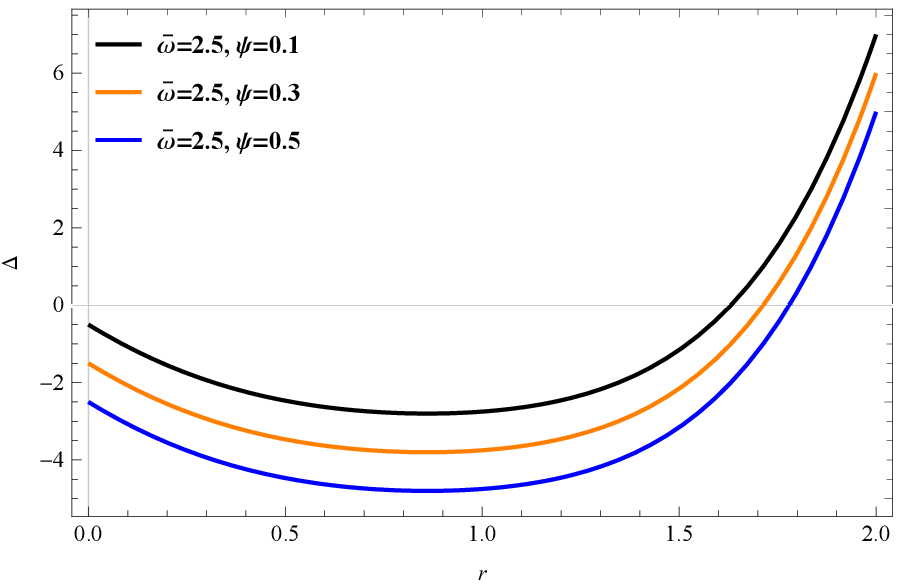}
\label{1a}}
\subfigure[]{\includegraphics[width=0.5\linewidth]{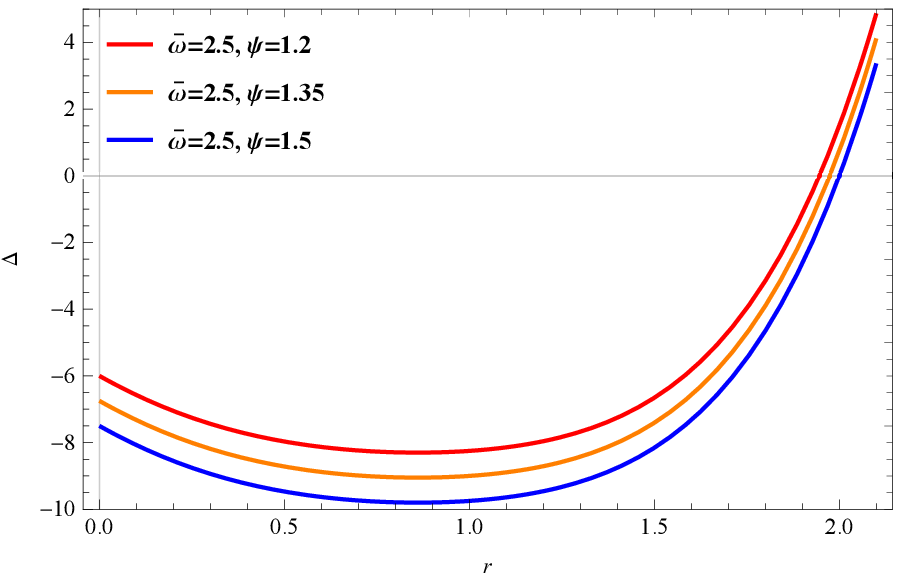}\label{1b}} 
\end{array}$
\end{center}
\caption{ In \ref{1a}, the behavior of horizon function $\Delta$ with respect to r by changing $\psi$=0.1,0.3 and 0.5. In \ref{1b}, the behavior of f(r) with respect to r by changing $\psi$=1.2,1.35 and 1.5. Here $\tilde\omega=\frac{118}{15\pi}$ and M=1.  }
\label{fig1}
\end{figure}

The positive real zeros of horizon function $\Delta$ gives the position of the horizon of this black hole. Figure \ref{fig1} illustrates the profile of an improved Schwarzschild horizon function $\Delta$ with respect to the radial coordinate r for three different values of free parameter $\psi$ with a fixed parameter $\tilde \omega =\frac{118}{15\pi}$. Left figure is for free parameter $\psi=0.1,0.3,0.5$ while the right figure is for varying free parameter $\psi=1.2,1.35,1.5$. Both plot shows the existence of one event horizon of the improved Schwarzschild  black hole, indicating one positive real root of the lapse function. To get the horizon we can consider only one constant value of fixed parameter whereas we can take different values of $\psi$ as mentioned above.

\section{Four-dimensional geodesics}\label{sec3}
Due to the spherical symmetry of an improved Schwarzschild  black hole, we can study the motion of the test particles on the equatorial plane, i.e at $\theta$ = $\frac{\pi}{2}$.  Therefoe, we have three geodesic equations:

\begin{equation}\label{7}
\frac{d^2t}{d\lambda^2}+\frac{B(r)}{A(r)}\frac{dt}{d\lambda}\frac{dr}{d\lambda}=0
\end{equation}

\begin{equation}\label{8}
\frac{d^2r}{d\lambda^2}+A(r)B(r)\Big(\frac{dt}{d\lambda}\Big)^2-\frac{B(r)}{A(r)}\Big(\frac{dr}{d\lambda}\Big)^2+rA(r)\Big(\frac{d\phi}{d\lambda}\Big)^2=0
\end{equation}

\begin{equation}\label{9}
\frac{d^2\phi}{d\lambda^2}+\frac{2}{r}\frac{dr}{d\lambda}\frac{d\psi}{d\lambda}=0
\end{equation}
where $A(r)=-f(r)$ and $B(r)$=$\frac{1}{r^2}\Big[-\frac{2\tilde\omega^2}{r^3}-\frac{5\tilde\omega^2\psi M}{r^4}-M+\frac{3M\tilde\omega}{r^2}\Big]$
\\
To solve this geodesic equations we will consider two approach in the next section. One is effective Newtonian potential approach as studied in \cite{g65} and another one is dynamical system analysis approach \cite{g66, g67}.
\section{Effective Newtonian potential approach}\label{sec4}
The lagrangian for particle motion for improved Schwarzschild black hole is given by
\begin{equation}\label{10}
2\mathcal{L}=-\Big(1-\frac{\tilde \omega^2}{r^4}-\frac{2\tilde \omega^2\psi M}{r^5}-\frac{2M}{r}+\frac{2M\tilde \omega}{r^3}\Big)\dot t^2+\frac{\dot r^2}{1-\frac{\tilde \omega^2}{r^4}-\frac{2\tilde \omega^2\psi M}{r^5}-\frac{2M}{r}+\frac{2M\tilde \omega}{r^3}}+r^2(\dot \theta^2+Sin^2\theta\dot\phi^2)
\end{equation}
here overdot denotes the differentiation with respect to $\lambda$ (affine parameter). Here lagrangian $\mathcal{L}$
is not depend explicitly on the coordinates t and $\phi$. Hence, due to this two cyclic coordinates we get two conserved
quantities, exemplary the energy (E) and the momentum (h) conjugate to $\phi$,
\\
\textbf{Energy}, 
\begin{equation}\label{11}
E=g_{tt}\frac{dt}{d\lambda}=-f(r)\frac{dt}{d\lambda}=A(r)\frac{dt}{d\lambda}
\end{equation}
\textbf{Momentum},
\begin{equation}\label{12}
2h=\frac{\partial{\mathcal{L}}}{\partial{\dot\phi}}=2r^2\dot\phi=constant
\end{equation}
here h indicates the total angular momentum of the particles. Recalling the normalization condition
\begin{equation}\label{13}
g_{\mu\nu}\frac{dx^{\mu}}{d\lambda}\frac{dx^{\nu}}{d\lambda}=-\epsilon
\end{equation}
for timelike geodesics $\epsilon$ = 1 and for null geodesics $\epsilon$ = 0 \cite{g33}. On the equatorial plane we have,
\begin{equation}\label{14}
\Big(\frac{dr}{d\lambda}\Big)^2=E^2-\frac{\Delta}{r^5}\Big(\frac{h^2}{r^2}+\epsilon\Big)=E^2+A(r)\Big(\frac{h^2}{r^2}+\epsilon\Big)
\end{equation}
Thus equations (\ref{11}), (\ref{12}) and (\ref{14}) are required to describe the dynamics of particle trajectories at the equatorial plane of an improved Schwarzschild black hole.
\\
Rewriting (\ref{14}),
\begin{equation}\label{15}
\frac{1}{2}\Big(\frac{dr}{d\lambda}\Big)^2=E_{eff}-V_{eff}
\end{equation}
where 
\begin{equation}\label{16}
E_{eff}=\frac{E^2}{2}
\end{equation}
and 
\begin{equation}\label{17}
V_{eff}=\frac{\Delta}{2r^5}\Big(\frac{h^2}{r^2}+\epsilon\Big)=\frac{1}{2}\Big(1-\frac{\tilde \omega^2}{r^4}-\frac{2\tilde \omega^2\psi M}{r^5}-\frac{2M}{r}+\frac{2M\tilde \omega}{r^3}\Big)\Big(\frac{h^2}{r^2}+\epsilon\Big)
\end{equation}
Therefore, Eq. (\ref{15}) represents the equation of motion of a particle having unit mass and an effective energy $E_{eff}$ moving in a one-dimensional(1D) effective potential $V_{eff}(r)$. Though E denotes the conserved energy of the particle per unit mass, $V_{eff}(r)$ for the radial coordinate r responds to $E_{eff}$ . This effective potential will be vanishes at the zeros of the lapse function. The physically acceptable regions are given by the values of those r for which $E_{eff}$ $>$ $V_{eff}(r)$ . Thus Eq. (\ref{15}) represents the energy equation for the radial
coordinate r. This equation is required to study the radial free fall and the stability of particle trajectories.
\\
By plugging A(r), equation (\ref{14}) takes the form:
\begin{equation}\label{18}
\Big(\frac{dr}{d\lambda}\Big)^2=E^2-\epsilon+\frac{2M\epsilon}{r}-\frac{h^2}{r^2}+\frac{2Mh^2}{r^3}-\frac{2M\tilde\omega \epsilon}{r^3}+\frac{\tilde\omega^2\epsilon}{r^4}+\frac{2\tilde\omega^2\psi M\epsilon}{r^5}-\frac{2M\tilde\omega h^2}{r^5}+\frac{\tilde\omega^2h^2}{r^6}+\frac{2\tilde\omega^2\psi Mh^2}{r^7}
\end{equation}
The geometry of the geodesics in the equatorial plane $\theta$ = $\frac{\pi}{2}$ can be determined by above Eq. (\ref{18}). To determine the shape of the trajectories we use (\ref{12}) to express $\frac{dr}{d\lambda}$ as
\begin{equation}\label{19}
\frac{dr}{d\lambda}=\frac{dr}{d\phi}\frac{d\phi}{d\lambda}=\frac{h}{r^2}\frac{dr}{d\phi}
\end{equation}
Now, introducing $u = \frac{1}{r}$, we arrive at
\begin{equation}\label{20}
\Big(\frac{du}{d\lambda}\Big)^2=u^4\Big(\frac{dr}{d\lambda}\Big)^2
\end{equation}
From Eq. (\ref{18})
\begin{eqnarray}\label{21}
\Big(\frac{du}{d\lambda}\Big)^2&=&2\tilde\omega^2\psi Mh^2u^{11}+\tilde\omega^2h^2u^{10}+(2\tilde\omega^2\psi M\epsilon-2M\tilde\omega h^2)u^9+\tilde\omega^2\epsilon u^8+(2Mh^2-2M\tilde\omega \epsilon)u^7\nonumber\\&-& h^2u^6+2M\epsilon u^5+(E^2-\epsilon)u^4
\end{eqnarray}
Now, using equations (\ref{19}), (\ref{20}) and (\ref{21}), we obtained the expression as following :
\begin{eqnarray}\label{22}
\Big(\frac{du}{d\phi}\Big)^2&=&2\tilde\omega^2\psi Mu^{7}+\tilde\omega^2u^{6}+\Big(\frac{2\tilde\omega^2\psi M\epsilon}{h^2}-2M\tilde\omega\Big)u^5+\frac{\tilde\omega^2\epsilon}{h^2} u^4+\Big(2M-\frac{2M\tilde\omega \epsilon}{h^2}\Big)u^3\nonumber\\&-& u^2+\frac{2M\epsilon}{h^2} u+\frac{E^2-\epsilon}{h^2}=S(u)
\end{eqnarray}
This equation describes the trajectories of the test particles near an improved Schwarzschild black hole. In the physical point of view, the radial
motion and the circular motion of the particles are two important aspects for studying the particles trajectories. Therefore, in next section we will discuss the particle's trajectories in these two cases.
\subsection{Radial motion}\label{sec5}
We need to study the radial geodesics for illustrating the essential features of the
space-time, for which angular momentum becomes zero. Also, for radial motion, $\phi$ = constant. Equation
(\ref{22}) fails to provide us the information about the radial trajectories as most of the terms blow up for h = 0. Moreover, to study the radial trajectories of the particles we can consider equation
(\ref{14}), which becomes
\begin{equation}\label{23}
\Big(\frac{dr}{d\lambda}\Big)^2=E^2+A(r)\epsilon
\end{equation}
\subsubsection{Motion of massive particles}\label{sec6}
For massive particles $\epsilon$ = 1. Eq. (\ref{23}) gives
\begin{equation}\label{24}
\Big(\frac{dr}{d\lambda}\Big)^2=E^2+A(r)=E^2-1+\frac{\tilde \omega^2}{r^4}+\frac{2\tilde \omega^2\psi M}{r^5}+\frac{2M}{r}-\frac{2M\tilde \omega}{r^3}
\end{equation}
By differentiating (\ref{24}) with respect to $\lambda$ and then dividing it by 2$\frac{dr}{d\lambda}$, we obtain
\begin{equation}\label{25}
\frac{d^2r}{d\lambda^2}=-\Big(\frac{2\tilde\omega^2}{r^5}+\frac{5\tilde\omega^2\psi M}{r^6}+\frac{M}{r^2}\Big)+\frac{3M\tilde\omega}{r^4}
\end{equation}
Since the trajectories of massive particles are characterized by timelike geodesics, so here we can assume the affine parameter proper time $\tau$ instead of $\lambda$ along the path. Hence the motion of massive particles can studied by the above equation. The condition for attractive
force per unit mass is
\begin{equation}\label{26}
\frac{2\tilde\omega^2}{r^5}+\frac{5\tilde\omega^2\psi M}{r^6}+\frac{M}{r^2}>\frac{3M\tilde\omega}{r^4}
\end{equation}
this condition is important to get the bound states of massive particles. The particle can acquire kinetic energy during the gravitational interaction when it goes through the gravitational field of any black hole. Then we can estimate the change in the potential energy in that gravitational field of the particle by taking into account particle's rest position such that the radial coordinate r changes to R. We can write Eq. (\ref{24}) in terms of the
affine parameter $\tau$, that gives
\begin{equation}\label{27}
\Big(\frac{dr}{d\tau}\Big)^2=(E^2-1)+\frac{\tilde \omega^2}{r^4}+\frac{2\tilde \omega^2\psi M}{r^5}+\frac{2M}{r}-\frac{2M\tilde \omega}{r^3}
\end{equation}
For r = R, $\frac{dr}{d\tau}$ = 0 and we can arrive at the equation for the kinetic energy per unit mass of the particle in terms of the change in its gravitational potential energy as follows:
\begin{equation}\label{28}
\frac{1}{2}\Big(\frac{dr}{d\tau}\Big)^2=\frac{\tilde \omega^2}{2}\Big(\frac{1}{r^4}-\frac{1}{R^4}\Big)+\tilde \omega^2\psi M\Big(\frac{1}{r^5}-\frac{1}{R^5}\Big)+M\Big(\frac{1}{r}-\frac{1}{R}\Big)-M\tilde \omega\Big(\frac{1}{r^3}-\frac{1}{R^3}\Big)
\end{equation}
Now, we are interested to represent the trajectories of the particle in the (t,r)-plane. With the help of Eqs. (\ref{11}) and (\ref{24}), 
\begin{equation}\label{28}
\Big(\frac{dr}{dt}\Big)^2=\Big(E^2+A(r)\Big)\frac{A^2(r)}{E^2}
\end{equation}
It is evident that $\frac{dr}{dt}$ = 0 for all points when $E^2$ = f(r) but f(r) $\ne$ 0 and E $\ne$ 0. From Eq. (\ref{24}) we get the effective potential $V_{eff}$ as follows :
\begin{equation}\label{29}
V_{eff}=\frac{1}{2}\Bigg(1-\frac{\tilde \omega^2}{r^4}-\frac{2\tilde \omega^2\psi M}{r^5}-\frac{2M}{r}+\frac{2M\tilde \omega}{r^3}\Bigg)
\end{equation}
\begin{figure}[h!]
\begin{center} 
$\begin{array}{cccc}
\subfigure[]{\includegraphics[width=0.35\linewidth]{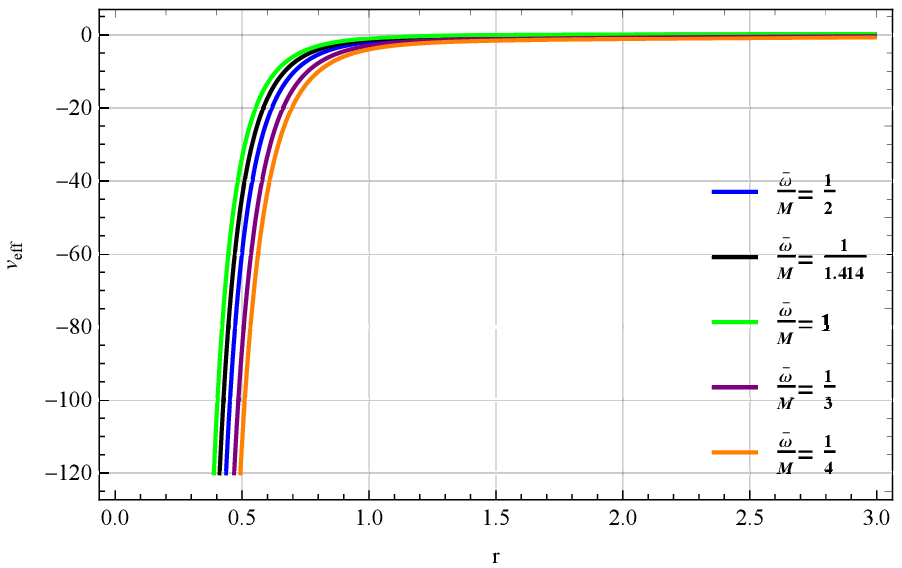}
\label{2a}}
\subfigure[]{\includegraphics[width=0.35\linewidth]{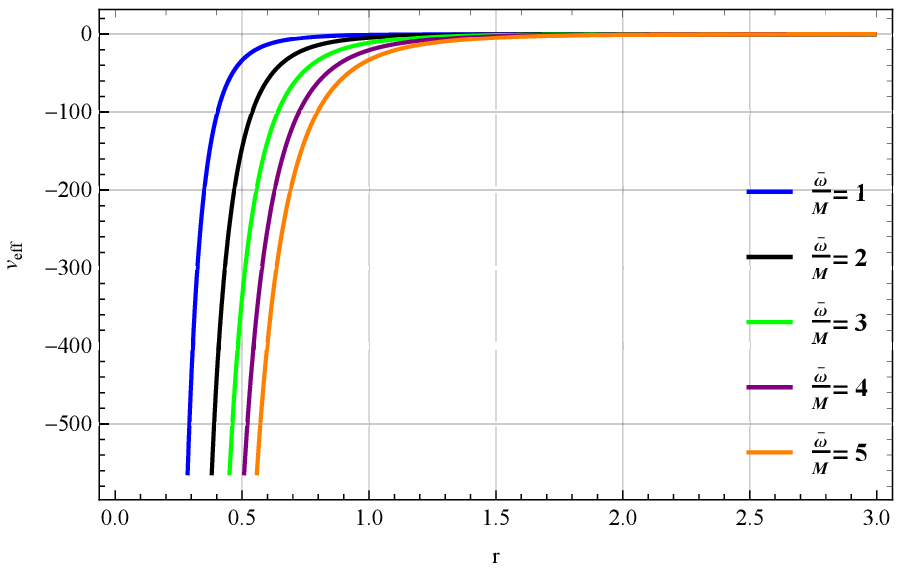}\label{2b}} 
\subfigure[]{\includegraphics[width=0.35\linewidth]{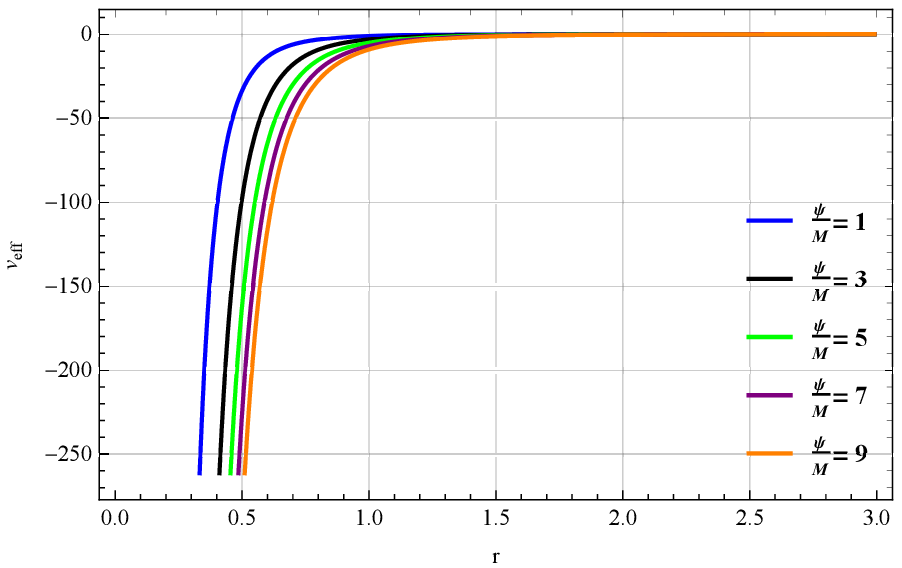}\label{2c}} 
\end{array}$
\end{center}
\caption{ In \ref{2a} and \ref{2b}, The behavior of $V_{eff}$ with respect to r by changing $\tilde\omega$ for a fixed $\psi$ = 1. In \ref{2c}, The behavior of $V_{eff}$ with respect to r by changing $\psi$ for a fixed $\tilde\omega$ = 1. Here, $M = 1$.  }
\label{fig2}
\end{figure}
Figure \ref{fig2} depicts the behavior of $V_{eff}$ with respect to the
radius r of the orbits for five different values of $\tilde \omega$ and $\psi$ of the black hole. From figure it is clear to us that, $V_{eff}$ is an exponentially increasing function of r but remains negative valued.
\\
Now, we can write from eq. (\ref{28})
\begin{equation}\label{30}
\frac{dr}{dt}=\frac{1}{E}\Bigg(1-\frac{\tilde \omega^2}{r^4}-\frac{2\tilde \omega^2\psi M}{r^5}-\frac{2M}{r}+\frac{2M\tilde \omega}{r^3}\Bigg)\Bigg(1-\frac{\tilde \omega^2}{r^4}-\frac{2\tilde \omega^2\psi M}{r^5}-\frac{2M}{r}+\frac{2M\tilde \omega}{r^3}\Bigg)^{\frac{1}{2}}
\end{equation}
The above equation won't be blow up for $r\rightarrow\infty$. Hence, this equation will only be a function of total energy E and therefore if one can observe the behavior of the particle at infinite distance from the source then obviously it will
behaves like a free particle. However, $\frac{dr}{dt}$ and E will be finite for a measurable distance from souce. So on (t,r)-plane, we can utilize (\ref{30}) in principle to get the correct trajectories.
\subsubsection{Motion of photons}\label{sec7}
For photons $\epsilon$ = 0. Eq. (\ref{23}) becomes
\begin{equation}\label{31}
\Big(\frac{dr}{d\lambda}\Big)^2=E^2
\end{equation}
where E is defined in eq. (\ref{11}). Also defining $\frac{dr}{dt}=\frac{dr}{d\lambda}\frac{d\lambda}{dt}$ we get,
\begin{equation}\label{32}
\frac{dr}{dt}=\pm A(r)=\mp\Bigg(1-\frac{\tilde \omega^2}{r^4}-\frac{2\tilde \omega^2\psi M}{r^5}-\frac{2M}{r}+\frac{2M\tilde \omega}{r^3}\Bigg)
\end{equation}
The solution of this equation will be complicated in nature and therefore it is difficult to describe the trajectories of the photons using this method. If we can consider a dynamical system by taking the variables $\frac{dr}{d\lambda}$ and $\frac{dt}{d\lambda}$, then Eq. (\ref{31}) will be able to gives us the information about the particle trajectories. We will show such an analysis later.
\subsection{Circular motion}\label{sec8}
It is clear from eq. (\ref{22}) that at equilibrium position of circular orbits $u=constant$, so $S(u) = 0$ and $S^{\prime}(u)$ = 0. From Eq. (\ref{22}) we have, 
\begin{eqnarray}\label{33}
S(u)&=&2\tilde\omega^2\psi Mu^{7}+\tilde\omega^2u^{6}+\Big(\frac{2\tilde\omega^2\psi M\epsilon}{h^2}-2M\tilde\omega\Big)u^5+\frac{\tilde\omega^2\epsilon}{h^2} u^4+\Big(2M-\frac{2M\tilde\omega \epsilon}{h^2}\Big)u^3\nonumber\\&-& u^2+\frac{2M\epsilon}{h^2} u+\frac{E^2-\epsilon}{h^2}
\end{eqnarray}
and 
\begin{eqnarray}\label{34}
S^{\prime}(u)&=&14\tilde\omega^2\psi Mu^{6}+6\tilde\omega^2u^{5}+5\Big(\frac{2\tilde\omega^2\psi M\epsilon}{h^2}-2M\tilde\omega\Big)u^4+\frac{4\tilde\omega^2\epsilon}{h^2}u^3+3\Big(2M-\frac{2M\tilde\omega \epsilon}{h^2}\Big)u^2\nonumber\\&-& 2u+\frac{2M\epsilon}{h^2}
\end{eqnarray}
After plugging the condition $S^{\prime}(u) = 0$ and $S(u) = 0$ in (\ref{34}), we get an expression of angular momentum (h) and energy (E) of the particle obeying circular motion, respectively, as follows:
\begin{eqnarray}\label{35}
h^2=\frac{\epsilon(5\tilde\omega^2\psi Mu^4+2\tilde\omega^2u^3-3M\tilde\omega u^2+M)}{u(1-7\tilde\omega^2\psi Mu^5-3\tilde\omega^2u^4+5M\tilde\omega u^3-3Mu)}
\end{eqnarray}
and
\begin{eqnarray}\label{36}
E^2=\frac{2\epsilon(1-2\tilde\omega^2\psi Mu^5-\tilde\omega^2u^4+2M\tilde\omega u^3-2Mu)}{(1-7\tilde\omega^2\psi Mu^5-3\tilde\omega^2u^4+5M\tilde\omega u^3-3Mu)}
\end{eqnarray}
\subsubsection{Motion of massive particles}\label{sec9}
For massive particles Eqs. (\ref{35}) and (\ref{36}) takes the form as, 
\begin{eqnarray}\label{37}
h^2=\frac{(5\tilde\omega^2\psi Mu^4+2\tilde\omega^2u^3-3M\tilde\omega u^2+M)}{u(1-7\tilde\omega^2\psi Mu^5-3\tilde\omega^2u^4+5M\tilde\omega u^3-3Mu)}
\end{eqnarray}
and
\begin{eqnarray}\label{38}
E^2=\frac{2(1-2\tilde\omega^2\psi Mu^5-\tilde\omega^2u^4+2M\tilde\omega u^3-2Mu)}{(1-7\tilde\omega^2\psi Mu^5-3\tilde\omega^2u^4+5M\tilde\omega u^3-3Mu)}
\end{eqnarray}
\begin{figure}[h!]
\begin{center} 
$\begin{array}{cccc}
\subfigure[]{\includegraphics[width=0.35\linewidth]{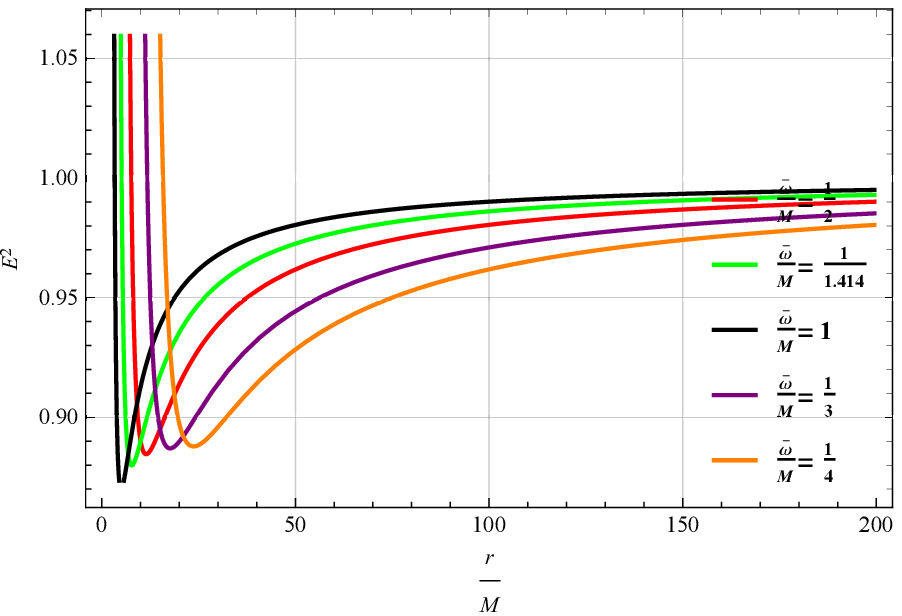}
\label{3a}}
\subfigure[]{\includegraphics[width=0.35\linewidth]{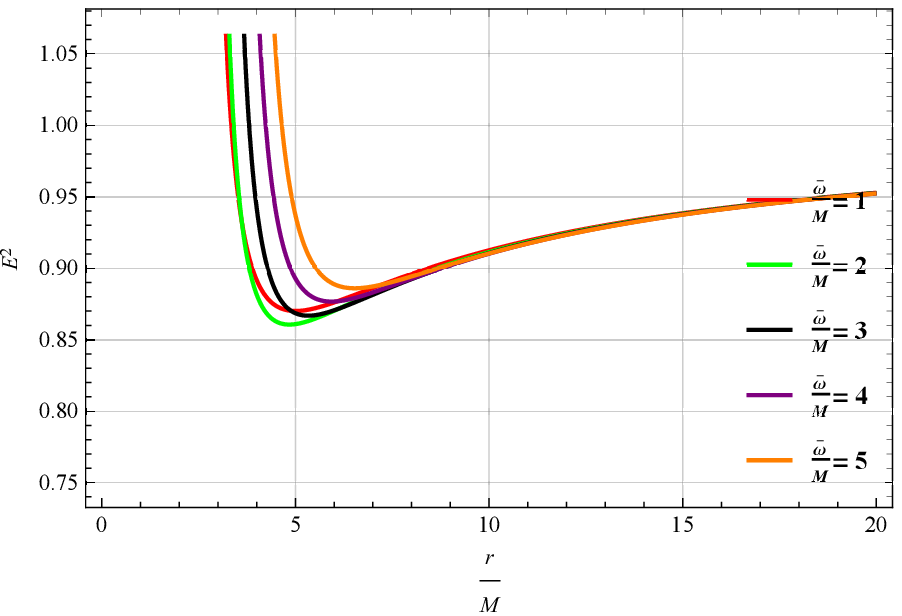}\label{3b}} 
\subfigure[]{\includegraphics[width=0.35\linewidth]{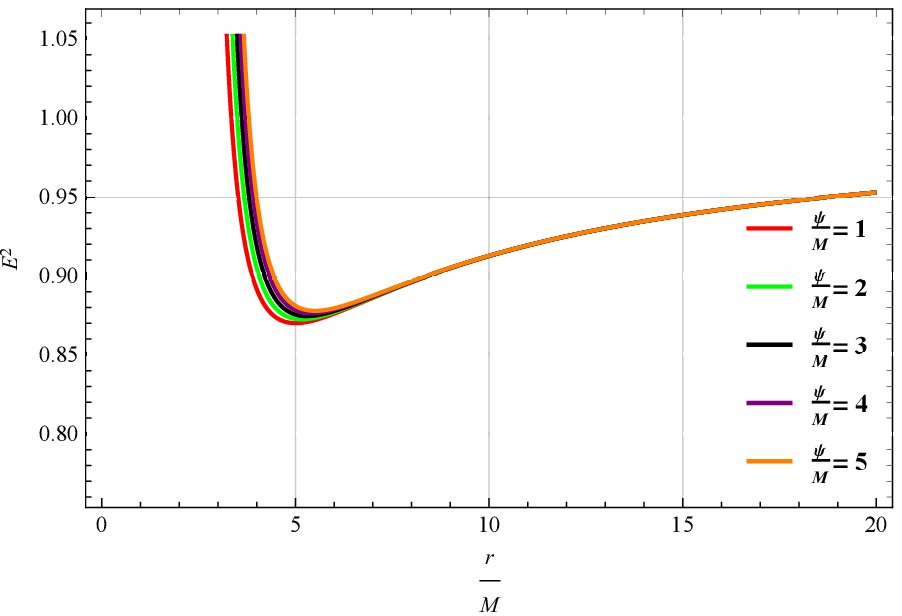}\label{3c}} 
\end{array}$
\end{center}
\caption{ In \ref{3a} and \ref{3b}, The behavior of $E^2$ with respect to $\frac{r}{M}$ by changing $\tilde\omega$ for a fixed $\psi$ = 1. In \ref{3c}, The behavior of $E^2$ with respect to $\frac{r}{M}$ by changing $\psi$ for a fixed $\tilde\omega$ = 1. Here, $M = 1$.  }
\label{fig3}
\end{figure}
The radius of the circular (bound) orbits can be calculated from eqs. (\ref{37}) and (\ref{38}). Moreover, the energy (E) of the particle
will be a constant of motion for radius r (=$\frac{1}{u}$) of a circular orbit. However, when the radius r varies E also vary. The behavior of $E^2$ vs specific radius $\frac{r}{M}$
for massive particles executing circular motion around an improved Schwarzschild black hole, for five different values of $\frac{\tilde\omega}{M}$ and $\frac{\psi}{M}$, is plotted in figure \ref{fig3}. Also, the variation of angular momentum $h^2$ with respect to specific radius $\frac{r}{M}$ for five different values of $\frac{\tilde\omega}{M}$ and $\frac{\psi}{M}$, is depicted in figure \ref{fig4}.
\begin{figure}[h!]
\begin{center} 
$\begin{array}{cccc}
\subfigure[]{\includegraphics[width=0.35\linewidth]{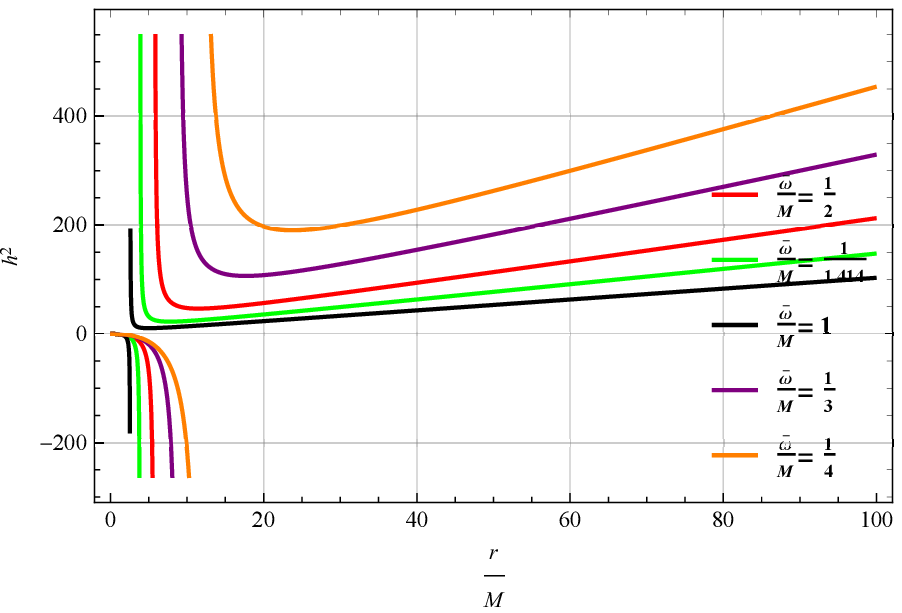}
\label{4a}}
\subfigure[]{\includegraphics[width=0.35\linewidth]{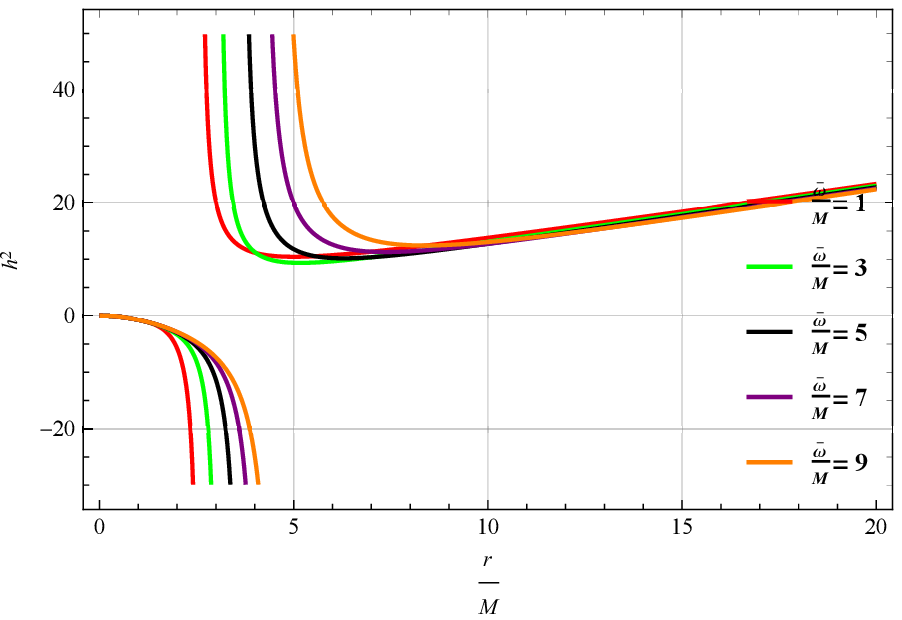}\label{4b}} 
\subfigure[]{\includegraphics[width=0.35\linewidth]{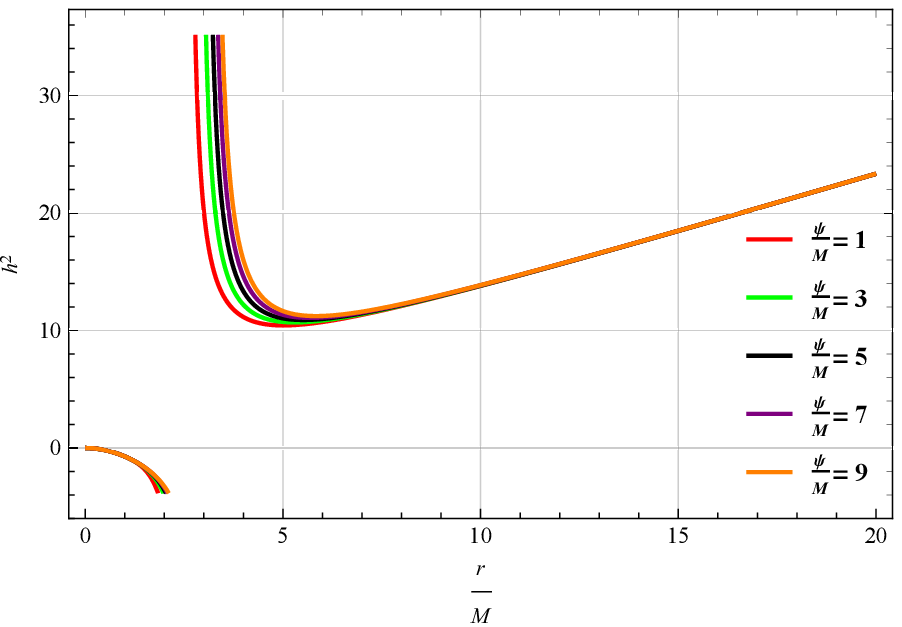}\label{4c}} 
\end{array}$
\end{center}
\caption{ In \ref{4a} and \ref{4b}, The behavior of $h^2$ with respect to $\frac{r}{M}$ by changing $\tilde\omega$ for a fixed $\psi$ = 1. In \ref{4c}, The behavior of $h^2$ with respect to $\frac{r}{M}$ by changing $\psi$ for a fixed $\tilde\omega$ = 1. Here, $M = 1$.  }
\label{fig4}
\end{figure}

In figures \ref{3a}, \ref{3b}, \ref{4a} and \ref{4b} the energy E and angular momentum h is a decreasing function of $\frac{r}{M}$ for different values of $\frac{\tilde\omega}{M}$ but constant free parameter $\psi$ while in figure \ref{3c} and \ref{4c}, both E and h is a decreasing function of $\frac{r}{M}$ for different values of $\frac{\psi}{M}$ but constant fixed parameter $\tilde\omega$. In figures \ref{3b} and \ref{4b}, energy and angular momentum increases for large values of $\frac{\tilde\omega}{M}$, indicating that for test particles to maintain in a stable circular orbit, a more energy E and angular momentum h for a larger values of $\frac{\tilde\omega}{M}$ is required. In figures \ref{3c} and \ref{4c}, energy and angular momentum increases for large values of $\frac{\psi}{M}$ respectively, indicating that for test particles to maintain in a stable
circular orbit, a more energy E and angular momentum h for a larger values of $\frac{\psi}{M}$ is needed. However, in \ref{3a} and \ref{4a}, for particles to maintain in a stable circular orbit, a more energy and angular momentum for a smaller values of $\frac{\tilde\omega}{M}$ is requisite. For increasing values of $\frac{r}{M}$, both E and h of all the plots are a increasing function. 
\subsubsection{Motion of photons}\label{sec10}
It is obvious that we can't calculate E and h of photons from eqs. (\ref{35}) and (\ref{36}). From (\ref{22}),
\begin{equation}\label{39}
\Big(\frac{du}{d\phi}\Big)=2\tilde\omega^2\psi Mu^7+\tilde\omega^2u^6-2M\tilde\omega u^5+2Mu^3-u^2+\frac{E}{h^2}=T(u)
\end{equation}
Hence, 
\begin{equation}\label{40}
T^{\prime}(u)=14\tilde\omega^2\psi Mu^6+6\tilde\omega^2u^5-10M\tilde\omega u^4+6Mu^2-2u
\end{equation}
For $T^{\prime}(u) = 0$, we get 
\begin{equation}\label{41}
7\tilde\omega^2\psi Mu^5+3\tilde\omega^2u^4-5M\tilde\omega u^3+3Mu-1=0
\end{equation}
It is a ploynomial equation of degree five, so it has five roots.
\\
For $T(u) = 0$, we obtain
\begin{equation}\label{42}
\frac{E}{h^2}=-2\tilde\omega^2\psi Mu^7-\tilde\omega^2u^6+2M\tilde\omega u^5-2Mu^3+u^2
\end{equation}
After putting one of the solution of (\ref{41}) in (\ref{42}), We can arrive at the relation between energy E and angular momentum h of photons in terms of the
geometric mass M of the improved Schwarzschild black hole, the free parameter $\psi$ and the fixed parameter $\tilde\omega$. 
\subsection{Stability of orbits for massive particles}\label{sec11}
\begin{figure}[h!]
\begin{center} 
$\begin{array}{cccc}
\subfigure[]{\includegraphics[width=0.35\linewidth]{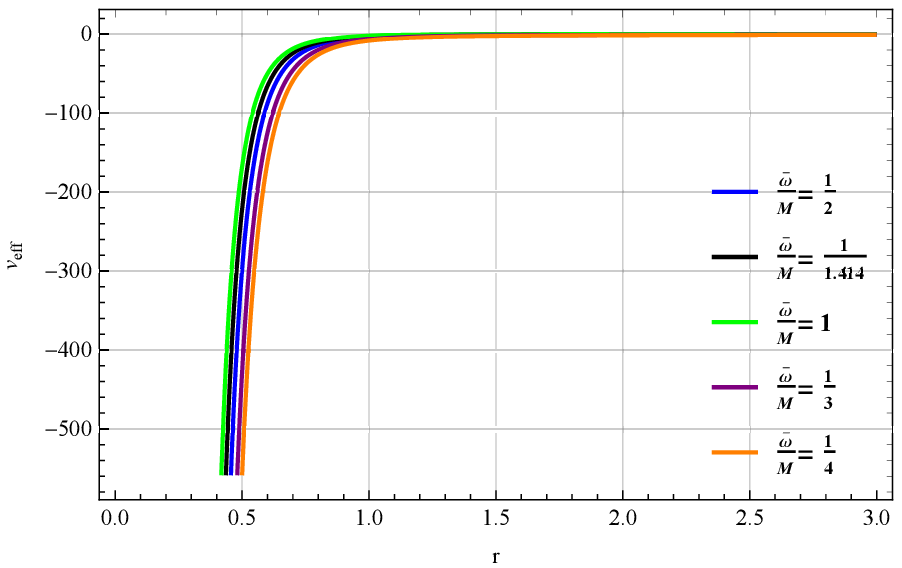}
\label{5a}}
\subfigure[]{\includegraphics[width=0.35\linewidth]{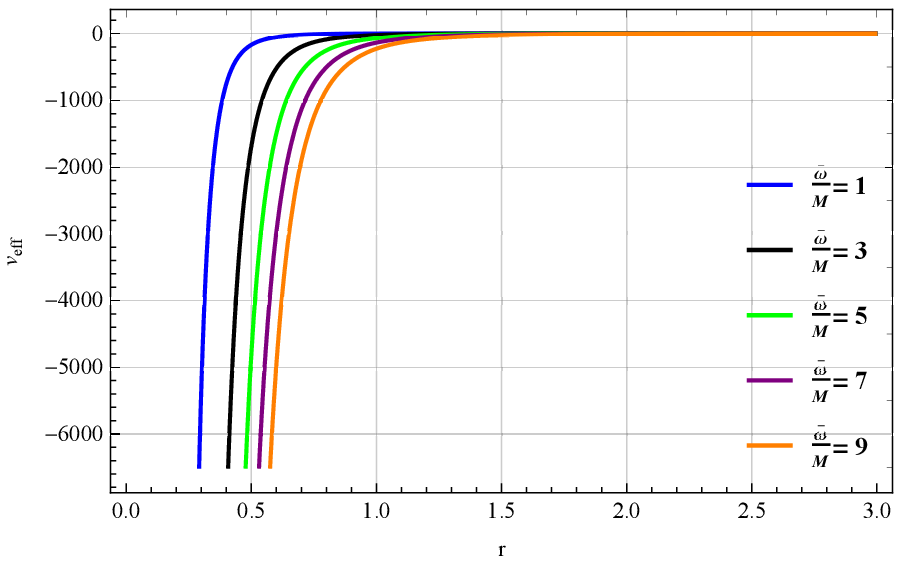}\label{5b}} 
\subfigure[]{\includegraphics[width=0.35\linewidth]{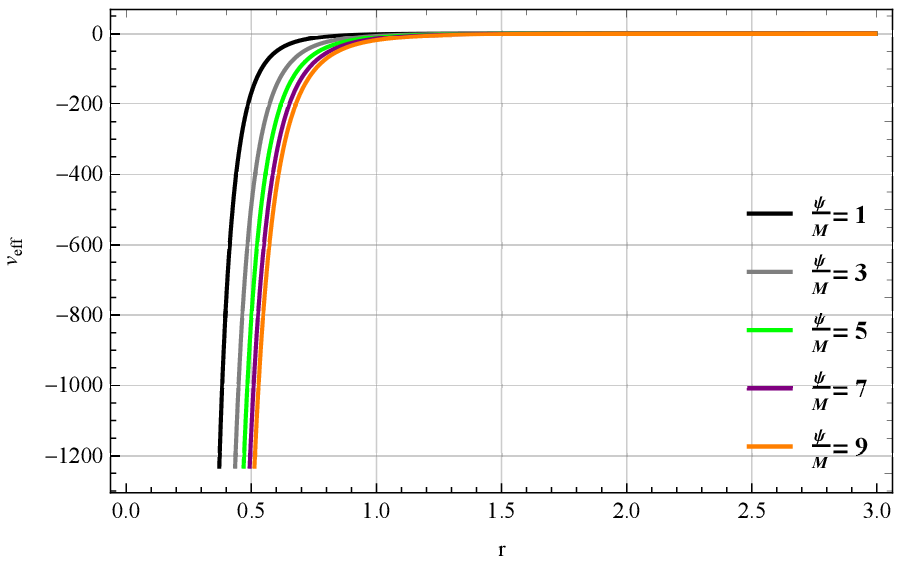}\label{5c}} 
\end{array}$
\end{center}
\caption{ In \ref{5a} and \ref{5b}, The behavior of $V_{eff}$ with respect to r by changing $\tilde\omega$ for a fixed $\psi$ = 1. In \ref{5c}, The behavior of $V_{eff}$ with respect to r by changing $\psi$ for a fixed $\tilde\omega$ = 1. Here, $M = 1$.  }
\label{fig5}
\end{figure}

We get freom (\ref{17}),
\begin{equation}\label{43}
V_{eff}=\frac{1}{2}\Bigg(1-\frac{\tilde \omega^2}{r^4}-\frac{2\tilde \omega^2\psi M}{r^5}-\frac{2M}{r}+\frac{2M\tilde \omega}{r^3}\Bigg)\Big(1+\frac{h^2}{r^2}\Big)
\end{equation}
The graph for effective potential vs $\frac{r}{M}$ for different values of $\frac{\tilde\omega}{M}$ and $\frac{\psi}{M}$, is shown in figure \ref{fig5}. The behavior of $V_{eff}$ is not similar to Newtonian potentials \cite{g28}. Here, we observe
that for all values of r, potential remains always negative which describes a bound system. From figure we see that effective potential graph has no maxima and minima. This clearly tell us that for each curve there exist only point of inflextion which indicates the existence of only marginally stable circular orbit ( known as innermost stable circular orbit, abbreviated as ISCO). Also the nature of this effective potential is different from the Schwarzschild black hole case \cite{g5}. If we can make $\tilde\omega = 0$ and $\psi = 0$ in (\ref{43}), it returns the well known Schwarzschild effective potential. The detailed study on effective potential and the probable orbits in Schwarzschild space-time has been discussed on \cite{g6, g7}.
\\
Stable circular orbits and unstable circular orbits gives the value of radius which represents the position of local minima and local maxima of the potential respectively. For stable circular orbits, we have $S(u) = 0$, $S^{\prime} = 0$ and $S^{\prime\prime} = 0$. We get from eq. (\ref{34})
\begin{equation}\label{44}
S^{\prime\prime}(u)=84\tilde\omega^2\psi Mu^5+30\tilde\omega^2u^4+40M\Big(\frac{\tilde\omega^2\psi }{h^2}-\tilde\omega\Big)u^3+\frac{12\tilde\omega^2}{h^2}u^2+12M\Big(1-\frac{\tilde\omega}{h^2}\Big)u-2
\end{equation}
For $S^{\prime\prime}(u) = 0$,
\begin{equation}\label{45}
84\tilde\omega^2\psi Mh^2u^5+30\tilde\omega^2h^2u^4+40M\Big(\tilde\omega^2\psi -\tilde\omega h^2\Big)u^3+12\tilde\omega^2u^2+12M\Big(h^2-\tilde\omega\Big)u-2h^2=0
\end{equation}
Therefore, for massive particles, to evaluate the innermost stable circular orbit (ISCO) near an improved Schwarzschild black hole we need to solve the polynomial of degree 5 in u. We can calculate the radius of ISCO of massive particles for improved Schwarzschild black hole from eq. (\ref{45}), provided $\tilde\omega\psi \ne h^2$ and $h^2\ne \tilde\omega$. An observer can see this track if this minimum radius will be bigger than the radius of event horizon.
\\
If $\psi = 0$, $\tilde\omega\psi  = h^2$ and $h^2 = \tilde\omega$ in eq. (\ref{45}), then ISCO of improved Schwarzschild black hole is given by,
\begin{equation}\label{46}
r^2_{ISCO}=\frac{60\tilde\omega^2h^2}{-12\tilde\omega\pm\sqrt{144\tilde\omega^2+240\tilde\omega^2h^4}}
\end{equation}
\subsection{Stability of orbits for photons}\label{sec12}
To discuss the stability of photon orbits, we get from (\ref{14})
\begin{equation}\label{47}
\Big(\frac{dr}{d\lambda}\Big)^2-A(r)\frac{h^2}{r^2}=E^2
\end{equation}
it takes the following form :
\begin{equation}\label{48}
\frac{1}{h^2}\Big(\frac{dr}{d\lambda}\Big)^2+\bar V_{eff}=\bar E_{eff}
\end{equation}
where $V_{eff} = \frac{1}{r^2}\Bigg(1-\frac{\tilde \omega^2}{r^4}-\frac{2\tilde \omega^2\psi M}{r^5}-\frac{2M}{r}+\frac{2M\tilde \omega}{r^3}\Bigg)$ and $E_{eff} = \frac{E^2}{L^2}$
\\
If the photon follows circular paths, then we estimate the minimum radius ($r_{mc}$) for the stable circular orbits of photon by considering the condition $T^{\prime\prime}(u) = 0$, which gives
\begin{equation}\label{49}
84\tilde\omega^2\psi Mu^5+30\tilde\omega^2u^4-40M\tilde\omega u^3+12Mu-2=0
\end{equation}
It is a ploynomial equation of degree five. Radius $r_{mc}$ of stable circular orbits can be estimated from (\ref{49}) if $\psi \ne 0$ and $\tilde\omega \ne o$. When free parameter $\psi$ of improved Schwarzschild black hole vanishes then (\ref{49}) reduces to a polynomial of degree four. Therefore $r_{mc}$ can be calculated as,
\begin{eqnarray}\label{50}
r_{mc}&>&\frac{3M}{2}+\frac{1}{2}\sqrt{9M^2+\frac{1202^{\frac{1}{3}}\tilde\omega(2M^2-\tilde\omega)}{\tilde\omega^2\mathcal{P}}+\frac{\mathcal{P}}{62^{\frac{1}{3}}\tilde\omega^2}}\\&+&\frac{1}{2}\sqrt{18M^2-\frac{1202^{\frac{1}{3}}\tilde\omega(2M^2-\tilde\omega)}{\tilde\omega^2\mathcal{P}}-\frac{\mathcal{P}}{62^{\frac{1}{3}}\tilde\omega^2}+\frac{216M^3-160M\tilde\omega}{\mathcal{D}}}
\end{eqnarray}
where, $\mathcal{P} = \Big(-30240M^2\tilde\omega^2-\sqrt{914457600M^4\tilde\omega^4-4(1440M^2\tilde\omega-720\tilde\omega^2)^3}\Big)^{\frac{1}{3}}$
\\
and $\mathcal{D} = 4\sqrt{9M^2+\frac{1202^{\frac{1}{3}}\tilde\omega(2M^2-\tilde\omega)}{\tilde\omega^2\mathcal{P}}+\frac{\mathcal{P}}{62^{\frac{1}{3}}\tilde\omega^2}}$
\section{ Dynamical systems analysis}\label{sec13}
In this section we will define geodesics equations in terms of dynamical variables and thereafter study its phase space. A detailed study on particle trajectories using dynamical system approach can be found in \cite{g8, g9}. 
Let us define three dynamical variable as follows :
\begin{equation}\label{51}
\frac{dt}{d\lambda}=X
\end{equation}
\begin{equation}\label{52}
\frac{dr}{d\lambda}=Y
\end{equation}
\begin{equation}\label{53}
\frac{d\phi}{d\lambda}=Z
\end{equation}
Now, geodesic eqs. (\ref{7}) - (\ref{9}) in terms of these three new variables,
\begin{equation}\label{54}
\frac{dX}{d\lambda}+\frac{B(r)}{A(r)}XY=0
\end{equation}
\begin{equation}\label{55}
\frac{dY}{d\lambda}+A(r)B(r)X^2-\frac{B(r)}{A(r)}Y^2+rA(r)W^2=0
\end{equation}
and
\begin{equation}\label{56}
\frac{dZ}{d\lambda}+\frac{1}{r}YZ=0
\end{equation}
X, Y and Z are ralated by the following equation
\begin{equation}\label{57}
A(r)X^2-\frac{1}{A(r)}Y^2+r^2Z^2=-\epsilon
\end{equation}
As W can be calculated from (\ref{57}), so studying of eqs. (\ref{54}), (\ref{55}) and (\ref{56}) is not necessary. Using Eqs. (\ref{57}) and (\ref{55}), we get
\begin{equation}\label{58}
\frac{dY}{d\lambda}+\frac{A(r)}{r}(rB(r)-A(r))X^2+\frac{1}{rA(r)}(A(r)-rB(r))Y^2-\frac{\epsilon A(r)}{r}=0
\end{equation}
\subsection{Real non-linear dynamical system}\label{sec14}
\begin{equation}\label{59}
\frac{dX}{d\lambda}=M(X,Y,r)
\end{equation}
\begin{equation}\label{60}
\frac{dY}{d\lambda}=N(X,Y,r)
\end{equation}
\begin{equation}\label{61}
\frac{dr}{d\lambda}=Y
\end{equation}
where $M(X,Y,r) = -\frac{B(r)}{A(r)XY}$
\\
$N(X,Y,r) = \frac{\epsilon A(r)}{r}-\frac{A(r)}{r}(rB(r)-A(r))X^2-\frac{1}{rA(r)}(A(r)-rB(r))Y^2 $
\subsection{Fixed point determination}\label{sec15}
To get the fixed point, we have to solve eqs. (\ref{59}), (\ref{60}) and (\ref{61}), for which Y, M and N becomes zero. Hence, equilibrium positions of particle trajectories can be obtained from these solutions. For phase space trajectories we consider the fixed point is $(X_{0},Y_{0} = 0, r_{0})$.
For a black hole with a given $\psi$, $\tilde\omega$ and mass M, the fixed point in the (X-Y) plane for null geodesics is (0, 0) whereas timelike geodesics produces definite fixed points. For timelike geodesics, 
\begin{equation}\label{62}
X_{0}=\pm\sqrt{\frac{1}{r_{0}B(r_{0}-A(r_{0}))}}
\end{equation}
\subsection{Phase space trajectories analysis}\label{sec16}
The phase evolution of the dynamical system in the (X, Y ) phase plane is obtained from the following equation :
\begin{equation}\label{63}
\frac{dY}{dX}=\frac{N}{M}
\end{equation}
here $M\ne 0$ and in case of null geodesics it gives the information of phase space evolution except at the point (0, 0).
\\
Plugging the values of M and N in (\ref{63}) and after integrating we obtain
\begin{equation}\label{64}
g_{1}X^2-g_{2}Y^2-g_{3}=0
\end{equation}
where, $g_{1} = \frac{A(r)C(r)}{3r}$, 
$g_{2} = \frac{C(r)}{rA(r)}+\frac{B(r)}{2A(r)}$, $g_{3} = \frac{\epsilon} A(r){r}$ and $C(r) = rB(r)-A(r)$. For null geodesics $g_{3} = 0$.
\begin{figure}[h!]
\begin{center} 
$\begin{array}{cccc}
\subfigure[]{\includegraphics[width=0.45\linewidth]{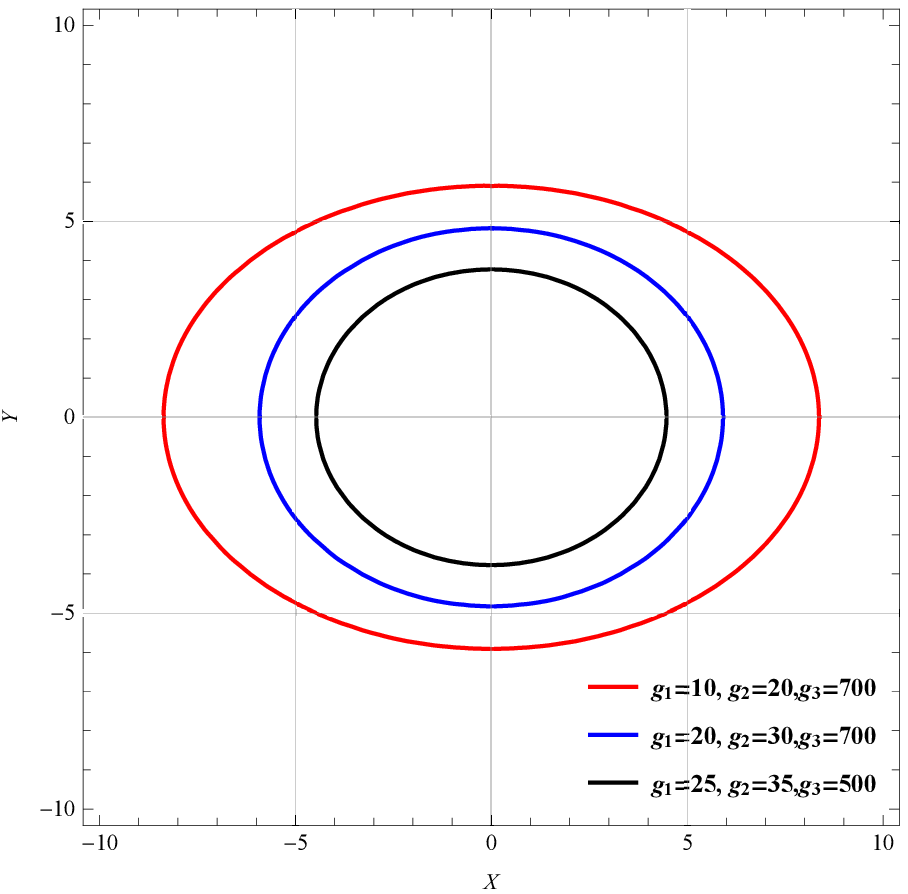}
\label{6a}}
\subfigure[]{\includegraphics[width=0.45\linewidth]{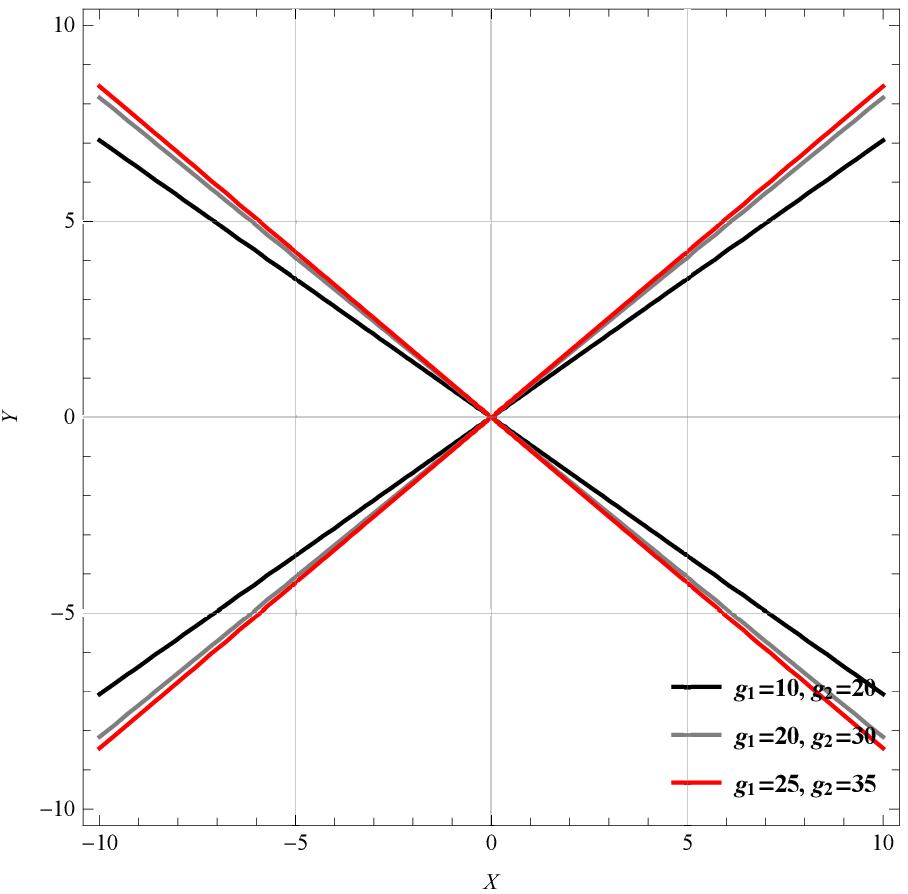}\label{6b}} 
\end{array}$
\end{center}
\caption{ In \ref{6a}, the plot for timelike geodesics for three sample equations $10X^2+20Y^2-700 = 0$, $20X^2+30Y^2-700 = 0$, 
$25X^2+35Y^2-500 = 0$. In \ref{6b}, the plot for null geodesics for three sample equations $-10X^2+20Y^2 = 0$, $-20X^2+30Y^2 = 0$, 
$-25X^2+35Y^2 = 0$.  }
\label{fig6}
\end{figure}

Plot \ref{6a} depicts the possible trajectories of timelike particles for different values of $g_{1}$, $g_{2}$ and $g_{3}$ on (X,Y)-plane. We see that the orbits are elliptical (periodic bpund) in nature. In plot \ref{6b}, the orbits of null like particles for different values of $g_{1}$ and $g_{2}$ has been shown. The phase trajectories of null particles are necessarily straight lines passing through the origin (0, 0), with their slopes get changes with the varying $g_{1}$ and $g_{2}$. 
\section{Conclusions and remarks}\label{sec17}
In this paper, We have investigated the timelike and null geodesics near an improved Schwarzschild black hole. We have determined the position of event horizon by plotting the lapse function of this black hole and the plot tells us that only one event horizon is physically acceptable. To investigate the circular motion and stability of circular orbits, we took inverse radial coordinate r and energy E and angular momentum h of massive particles has been calculated. The nature of the  particle trajectories depend upon the energy, angular momentum, free parameter $\psi$ and fixed parameter $\tilde{\omega}$. We have noticed that the enargy and angular momentum of massive particles is a decreasing function of specific radius for different values of $\frac{\tilde\omega}{M}$ and $\frac{\psi}{M}$ but for increasing values of specific radius both energy and angular momentum are increasing function. We have seen that the radius of the innermost stable circular
orbit (ISCO) of massive particles is completely defined in terms of their angular momentum h and fixed parameter $\tilde{\omega}$, free parameter $\psi$ and mass M of the black hole. Since it is impossible to obtain the possible trajectories using Newtonian potential method, so simultaneously we have studied dynamical systems. Also, We have noticed
that the geodesics of timelike (massive) particles near an improved Schwaezschild black hole possesses only periodic bound orbits and for these orbits there are definite fixed point. Moreover, the null geodesics produces a unique fixed point and these trajectories are terminating orbits in nature.
\\
\\
\textbf{Data availibility}
\\
 Data sharing is not applicable to this article, as no data sets were generated or analyzed during the current study.
 \\
 \\
 \textbf{Acknowledgements}
 \\
The author is thankful to Dr. Sudhaker Upadhyay, department of physics, K. L. S. College, Nawada, Bihar 805110, India, for various suggestions which developed the presentation of the paper.
 \\

\end{document}